\documentclass[pra,aps,superscriptaddress,showpacs]{revtex4}
\usepackage{epsfig}
\usepackage{graphicx}
\usepackage{amsmath}
\usepackage{amscd}

\newcommand{\beq}{\begin{eqnarray}}
\newcommand{\eeq}{\end{eqnarray}}
\def\be{\begin{equation}}
\def\ee{\end{equation}}
\def\ba{\begin{eqnarray}}
\def\ea{\end{eqnarray}}

\begin{document}

\title
{Driven Dirac-like Equation via Mirror Oscillation: Controlled Cold-Atom Zitterbewegung}
\author{Qi Zhang}
\affiliation{Centre of Quantum Technologies and Department of
Physics, National University of Singapore, 117543, Singapore}

\author{Jiangbin Gong}
\affiliation{Department of Physics and Centre for Computational
Science and Engineering, National University of Singapore, 117542,
Singapore}
 \affiliation{NUS Graduate School for Integrative Sciences
and Engineering, Singapore 117597, Republic of Singapore}

\author{C.H. Oh}
\affiliation{Centre of Quantum Technologies and Department of
Physics, National University of Singapore, 117543, Singapore}
\affiliation{Institute of Advanced Studies, Nanyang Technological
University, Singapore 639798, Republic of Singapore}

\date{\today}
%
\date{\today}
\begin{abstract}
By considering mirror oscillation in a ``tripod-scheme" laser-atom system, we advocate explorative studies of driven Dirac-like equations. Both analytical and numerical studies show that mirror oscillation can be used to drive an effective spin-orbit interaction and hence control the amplitude, the frequency, and the damping of the cold-atom Zitterbewegung oscillation.  Our results demonstrate an interesting coupling between the mirror mechanical motion and a fundamental quantum coherent oscillation, opening up new means of matter wave manipulation.
\end{abstract}
\maketitle

\section{introduction}
The jittering motion of a free relativistic electron predicted by the
Dirac equation, called Zitterbewegung oscillation (ZB)
\cite{Schrodinger}, is a truly fundamental quantum coherence effect.
Directly observing the free-electron ZB is, however, practically impossible
due to its extremely high frequency and small amplitude. For this
very reason, studies of ZB effect necessarily call for quantum
simulations based on Dirac-like equations with (effective)
spin-orbit coupling, including those involving band electrons in
graphene \cite{Novoselov}, cavity electrodynamics \cite{larson}, single trapped ion \cite{ion1,ion2},
as well as ultracold atoms \cite{JuzeliunasPRA2008,VaishnavPRL2008,MerklEPL2008}.

To actively
explore ZB-related physics, it is necessary to go beyond passive simulations of the known ZB effect.
Here we advocate to consider Dirac-like
equations driven by an external field.  This is feasible due to the
precise controllability of laser-atom interaction. Given the current
vast interest in dressed matter waves \cite{dressed}, this topic is
also timely because it can reveal how a driven effective spin-orbit
interaction may be used as a new means of matter-wave manipulation.
Furthermore, as proposed below, a driven Dirac-like equation can be
achieved via mirror oscillation, thus directly coupling the
fascinating cold-atom ZB with the mirror mechanical motion. This
interesting quantum-classical interface may lead to a novel setup of
optomechanical systems \cite{opo}, with spin-orbit interaction also included.

In particular, to realize a driven Dirac-like equation we propose to
add oscillating mirrors [see Fig. 1(b)] to a recent cold-atom-ZB
scheme \cite{VaishnavPRL2008} that involves tripod-scheme cold atoms
\cite{Bergmann,Ruseckas2005PRL,Juzeliunas2008PRL} interacting with
three laser fields.  This modification induces a coupling between
the mirror mechanical motion and the cold-atom matter wave, yielding
a time-dependent effective spin-orbit interaction. It is shown that
the amplitude of the cold-atom ZB can then be either enhanced or
weakened. Such control over the ZB amplitude is also related to a
dynamical realization of an effective ``spin-helix" Hamiltonian
\cite{zhang} and the celebrated phenomenon of ``coherent destruction
of tunneling" (CDT) \cite{CDT} in driven systems. More remarkably,
the quick damping of the cold-atom ZB, which hinders experimental
studies, can also be dramatically suppressed. Extending the lifetime
of cold-atom ZB might be useful for finding its applications in
precision measurements or sensing.  It should be also stressed that although we
present our findings in the cold-atom context throughout this paper, they should have
direct analogs in other alternative ZB realizations mentioned above.


\section{DRIVEN DIRAC-LIKE EQUATION}
Consider then the interaction of a tripod-scheme cold-atom
interacting with three laser fields \cite{Bergmann,Ruseckas2005PRL}
[see Fig.1 (a)]. The four internal atomic levels are denoted
$|n\rangle$, with $n=0-3$, with the three states $|1\rangle$,
$|2\rangle$, and $|3\rangle$ being degenerate magnetic sub-levels on
the ground state. Each of the three laser fields has an appropriate
polarization and induces a transition $|0\rangle \leftrightarrow
|n\rangle$, with the Rabi frequency $\Omega_{n}$, $n=1-3$. In the
interaction picture and under the rotating wave approximation the
internal Hamiltonian is given by
$H_{\text{RWA},4}=\hbar\sum_{n=1}^{3}(\Omega_{n}|0\rangle\langle n|
+ \Omega_{n}^{*}|n\rangle\langle 0|)$.  This Hamiltonian possesses
two dark states $|D_1\rangle$, $|D_2\rangle$, with zero eigenvalue
and zero overlap with the excited state $|0\rangle$. For
sufficiently large Rabi frequencies $\Omega_{n}$ and for slow
translation motion (for the system parameters associated with all our computational examples below,
it can estimated that the characteristic Rabi frequency should be around $10^7$ Hz or higher), the internal state can remain in a
two-dimensional dark subspace spanned by $|D_1\rangle$ and
$|D_2\rangle$. In the dark state representation which can be
space-dependent, the translation motion of the atom effectively
experiences a non-Abelian gauge field. The total stationary Hamiltonian
then becomes
\begin{equation} \label{original}
H_{S}=\frac{1}{2m}(\mathbf{p}-\mathbf{A})^{2}+\Phi+V,
\end{equation}
where $m$ is atom mass, ${\bf p}$ is atom momentum,
$\mathbf{A}_{ij}=i\hbar\langle D_{i}|\nabla D_{j}\rangle$ is the
effective non-Abelian $2\times 2$ potential matrix, $\Phi$ is the
Born-Huang scalar potential, and $V$ is the external potential.
For convenience we assume $\Phi+V=\text{constant}$
and adopt the laser configuration studied in Ref.
\cite{{Juzeliunas2008PRL}}, with two laser fields
counter-propagating along $x$, and another propagating along $z$,
$\Omega_{1}=\Omega_{0}\sin(\xi)e^{-ik_lx}/\sqrt{2}$,
$\Omega_{2}=\Omega_{0}\sin(\xi)e^{ik_lx}/\sqrt{2}$,
$\Omega_{3}=\Omega_{0}\cos(\xi)e^{ik_lz}$, and
$\cos(\xi)=\sqrt{2}-1$, where $k_l$ is the laser wavevector. It then
suffices to consider the $x-z$ plane in Fig. 1(b) with two
unit vectors $\hat{e}_x$ and $\hat{e}_z$. As such, ${\bf p}=p_x \hat{e}_x+
p_z\hat{e}_z$, and the Hamiltonian $H_{S}$ becomes

\begin{figure}[t]
\begin{center}
\vspace*{-0.5cm}
\par
\resizebox *{11cm}{7.6cm}{\includegraphics*{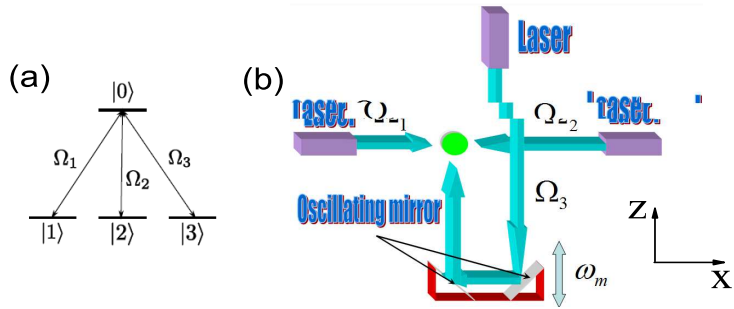}}
\end{center}
\par
\vspace*{-1.0cm} \caption{(Color Online) (a) Level structure of
tripod-scheme cold atoms. (b) Schematic setup involving oscillating
mirrors and three laser fields interacting with cold atoms. As shown
in the text, this leads to a driven Dirac-like equation.}
\end{figure}

\begin{equation} \label{DiracH}
H_{S}=\frac{{\bf p}^{2}}{2m}+\frac{\hbar
\kappa}{m}(p_{x}\sigma_{x}+p_{z}\sigma_{z}),
\end{equation}
where $\kappa\equiv (\sqrt{2}-1) k_{l}$,  and $\sigma_{x,z}$ are the
standard Pauli matrices. The second term in Eq. (\ref{DiracH})
represents an effective spin-orbit coupling that resembles the
Dresselhaus spin-orbit Hamiltonian (other types of spin-orbit
interaction, such as a mixture of Dresselhaus and Rashba coupling,
can be obtained by considering different laser configurations \cite{VaishnavPRL2008}). When the effects of
the ${\bf p}^2/2m$ term are small as compared with the spin-orbit coupling term,
the Schr\"odinger equation for $H_S$ will be dominated by
the second spin-orbit term in Eq. (\ref{DiracH}), which is linear in $p_x$ or $p_z$.  In this sense
a Dirac-like equation (with two-component spinors) is obtained.

To arrive at a driven Dirac-like equation, we consider two oscillating
mirrors joined together, schematically shown in Fig. 1(b). If the
mirrors are moving at a velocity $v/2=(v_{d}/2)\cos(\omega_d t)$ in
the $z$ direction, then effectively the $\Omega_3$ field is moving
towards the atom at a velocity $v$. As indicated below, such a
movement is very slow as compared with the speed of light and hence
relativistic Doppler effects can be safely neglected. In a moving
frame where the laser source is stationary, Eq. (\ref{DiracH})
applies.  We then make a Galileo transformation back to the
laboratory frame, obtaining the following driven Hamiltonian
\begin{equation} \label{DiracHtransz}
H_{\text{driven}}=\frac{{\bf p}^{2}}{2m}+\frac{\hbar
\kappa}{m}\{[p_{z}-mv_{d}\cos(\omega_{d}t)]\sigma_{z}+p_{x}\sigma_{x}\}.
\end{equation}
This driven Hamiltonian can be alternatively derived by considering
non-Abelian geometric phases induced by the moving mirrors
\cite{Zhang2009PRA}. When the spin-orbit interaction in Eq.
(\ref{DiracHtransz}) (which is now time-dependent) accounts for the
main physics, the Schr\"{o}dinger equation for $H_{\text{driven}}$
simulates a driven Dirac-like equation.  In the following two driven
cases will be elaborated as examples.

\section{CONTROLLED COLD-ATOM ZITTERBEWEGUNG}
Figure 2 depicts the time-dependence of the expectation value
$\langle x\rangle$ or $\langle z\rangle $ of the cold atom,
calculated from the evolution associated with $H_{\text{driven}}$.
The initial state is given by
$|\Psi(0)\rangle=(1/\sqrt{2},i/\sqrt{2})^{\text{T}} \int d{\bf p}
|{\bf p}\rangle G({\bf p})$, where $\langle {\bf r}|{\bf p}\rangle
\sim e^{i{\bf p}\cdot {\bf r}}$, and $G({\bf p})$ is a Gaussian
distribution. This initial state represents a two-component spinor
in the dark state representation times a Gaussian wavepacket. The
shown oscillations in $\langle x\rangle$ or $\langle z\rangle$ without mirror oscillation
(solid lines) are the expected cold-atom ZB effect. But remarkably, the ZB amplitude
and frequency are seen to be strongly affected by introducing the
mirror oscillation. In particular, in the case of Fig. 2(a), the
initial wavepacket is set to move along the $z$ direction, ZB occurs
in $\langle x\rangle$, with its amplitude tuned down extensively and
with its frequency unchanged. In the case of Fig. 2(b), the initial
wavepacket is set to move along the $x$ direction, ZB occurs in
$\langle z\rangle $, with its amplitude tuned up extensively and at
the same time its frequency much decreased. Note that in all the
shown examples the ZB frequency is much smaller than the mirror
oscillation frequency $\omega_{d}$.

\begin{figure}[t]
\begin{center}
\vspace*{-0.cm}
\par
\resizebox *{12cm}{10.4cm}{\includegraphics*{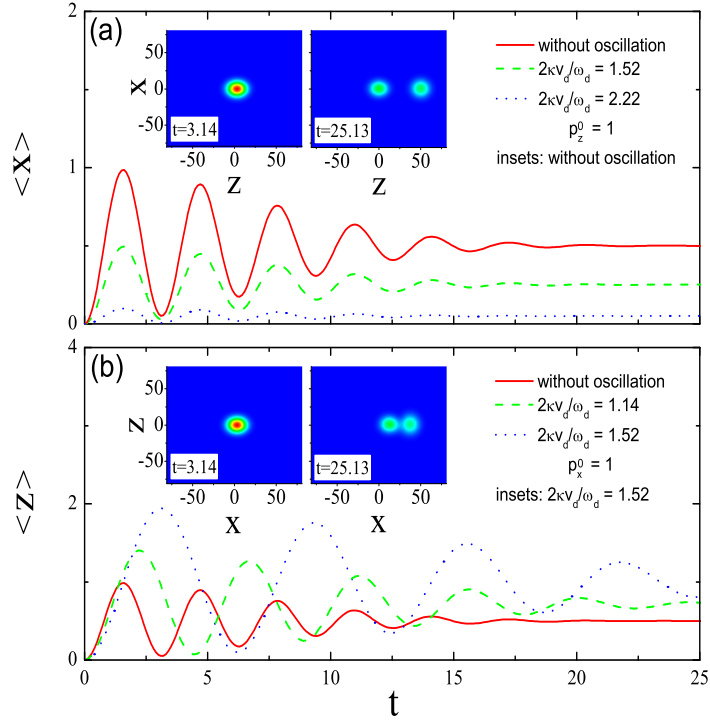}}
\end{center}
\par
\vspace*{-0.5cm} \caption{(Color Online) Effects of a high-frequency
mirror oscillation on cold-atom ZB. The initial momentum
$\mathbf{p}_0$ is taken in the $z$ direction in (a) and in the $x$
direction in (b). Insets are the typical wavepacket profile at early
and later times, with the initial variance in $x$ and $z$ being around 10.0.
 Throughout $t$ is in units of $m/(\hbar
\kappa^{2})$, $x$ and $z$ are in units of $1/\kappa$.
$\omega_{d}=50$ in units of $\hbar\kappa^{2}/m$. }
\end{figure}

To understand the results in Fig. 2, we adopt a perspective afforded
by a standard high-frequency approximation. That is, when $1/\omega_d$
is much smaller than all other time scales, the driven dynamics can
be approximately described by a static effective Hamiltonian
$H^{\text{eff}}_{\text{driven}}$ via an averaging of the
time-dependent term of $H_{\text{driven}}$, i.e., $-\hbar \kappa v_d
\cos(\omega_{d}t)\sigma_z =H_{\text{driven}}-H_S\equiv A(t)$. To the
first order of $1/\omega_{d}$, we obtain
$H^{\text{eff}}_{\text{driven}}=
(\omega_{d}/2\pi)\int_{0}^{2\pi/\omega_{d}}dt \ e^{iF(t)} H_{S}
e^{-iF(t)}$, where $F(t)=\int_{0}^{t} dt' A(t')/\hbar $
\cite{Hanggi-review}. Specifically,
\begin{equation} \label{Bessel}
H^{\text{eff}}_{\text{driven}}=\frac{{\bf p}^{2}}{2m}+\frac{\hbar
\kappa}{m}[p_{z}\sigma_{z}+{\cal J}_{0}(\frac{2\kappa
v_{d}}{\omega_{d}})p_{x}\sigma_{x}],
\end{equation}
where ${\cal J}_0$ is the ordinary Bessel function of order zero. Let
$\theta(\tilde{{\bf p}})$ be the angle between the vector
$\tilde{{\bf p}}\equiv p_x {\cal J}_0(\frac{2\kappa
v_{d}}{\omega_{d}}) \hat{e}_{x} + p_z \hat{e}_z$ and the $x$
axis, and let $\beta(\tilde{{\bf p}})=\pi/4-\theta(\tilde{{\bf
p}})/2$. The eigenstates of $H^{\text{eff}}_{\text{driven}}$ are
then found to be $|\psi^{+}\rangle = [\cos(\beta),\sin(\beta)]^{\text{T}} |{\bf
p}\rangle $ and $|\psi^{-}\rangle =
[\sin(\beta),-\cos(\beta)]^{\text{T}} |{\bf p}\rangle$, with their eigenvalues given by $ E_{\pm}({\bf p}, \tilde{\bf
p})\equiv {\bf p}^2/2m \pm \hbar \kappa |\tilde{{\bf p}}|/m$.
Because each ${\bf p}$ component in the initial state
$|\Psi(0)\rangle$  can be expanded by the eigenstates
$|\psi^{\pm}\rangle$, it can be predicted from
$H^{\text{eff}}_{\text{driven}}$ that the total state
$|\Psi(t)\rangle$ at time $t$ is given by
\begin{eqnarray}
\label{psit} |\Psi(t)\rangle &=&\frac{1}{\sqrt{2}} \int d{\bf p} G({\bf p})
e^{i\beta(\tilde{{\bf p}})}
\nonumber \\
&\times &
\left[
e^{-\frac{iE_{+}({\bf p}, \tilde{\bf p})t}{\hbar}}|\psi^{+}\rangle-i e^{-\frac{iE_{-}({\bf p}, \tilde{\bf p})t}{\hbar}}|\psi^{-}\rangle \right].
\end{eqnarray}
Further using ${\bf r}=i\hbar {\bf \nabla}_{\bf p}$, one easily finds
\begin{eqnarray}
\langle \mathbf{r}(t)\rangle&=&i\hbar\langle
\Psi(t)|\nabla_{\mathbf{p}}|\Psi(t)\rangle  = {\bf r}^0 + {\bf
p}^{0}t/m \nonumber \\ &+&\frac{\hbar}{2}\int d\mathbf{p}
|G(\mathbf{p})|^{2}[\nabla_{\mathbf{p}}\theta(\tilde{\mathbf{p}})]\{1-\cos[\omega(\tilde{\mathbf{p}})t]\},
\label{com}
\end{eqnarray} where the last oscillating term
represents cold-atom ZB, with the angular frequency
$\omega(\tilde{\mathbf{p}})\equiv
[E_{+}({\bf p},\tilde{\mathbf{p}})-E_{-}({\bf p},\tilde{\mathbf{p}})]/\hbar = 2\kappa |\tilde{\bf p}|/m $,
 ${\bf r}^0$ is the initial expectation value of ${\bf r}$, and ${\bf p}^0$ is the initial momentum of the Gaussian wavepacket.

We now come back to the results in Fig. 2.  In the case of Fig.
2(a), $p_x\approx 0$, $\mathbf{p}\approx p_z\hat{e}_z$, then
$\omega(\tilde{\mathbf{p}})\approx \omega({\bf p})$ and
\begin{equation}
\nabla_{\mathbf{p}}\theta(\mathbf{\tilde{p}})\approx -\frac{{\cal
J}_0(\frac{2\kappa v_{d}}{\omega_d})}{p_z}\hat{e}_{x}.
\label{case1}
\end{equation}
Substituting these two relations into Eq. (\ref{com}), one directly
obtains that ZB is along $x$, its
amplitude is proportional to the factor ${\cal J}_0(\frac{2\kappa
v_{d}}{\omega_d})$, and its frequency is independent of
$\omega_{d}$. Quantitatively, for the two shown examples with
$2\kappa v_{d}/\omega_d = 1.52$ or 2.22 in Fig. 2(a), the ZB
amplitude should decrease by a factor of 2.0 or 10.0 as compared
with that without mirror oscillation ($v_{d}=0)$, in excellent
agreement with the numerics.  Likewise, in the case of Fig. 2(b),
$p_z\approx 0$, $\mathbf{p} \approx p_x\hat{e}_x$, leading to $\omega
(\tilde{\mathbf{p}})\approx {\cal J}_0(\frac{2\kappa
v_{d}}{\omega_d}) \omega ({\bf p})$ and
\begin{equation}
\nabla_{\mathbf{p}}\theta(\mathbf{\tilde{p}})\approx \frac{1}{p_x{\cal
}J_0(\frac{2\kappa v_{d}}{\omega_d})}\hat{e}_{z}.
\end{equation}
Equation (\ref{com}) then predicts that ZB is now
along $z$, its amplitude is proportional to the factor $1/{\cal
J}_0(\frac{2\kappa v_{d}}{\omega_d})$, and the ZB frequency is
proportional to ${\cal J}_0(\frac{2\kappa v_{d}}{\omega_d})$. For
the two examples with $2\kappa v_{d}/\omega_d = 1.14$ or 1.52 in
Fig. 2(b), the ZB amplitude should be enhanced by a factor of 1.41
or 2.0 as compared with that without mirror oscillation, and the
associated ZB frequency should be decreased by a factor 0.7 or 0.5.
This is again in agreement with our direct numerical experiments.
Note that if we have ${\cal J}_0(\frac{2\kappa v_{d}}{\omega_d})=0$,
i.e., exactly on a CDT point \cite{CDT} where the transition between
the two dark states is totally suppressed by the driving, then in
the case of Fig. 2(a), the ZB amplitude is zero, and in the case of
Fig. 2(b), the ZB frequency is zero. So in either case the ZB effect
disappears on a CDT point.  Interestingly, this is also a situation
where $H^{\text{eff}}_{\text{driven}}$ takes a spin-helix
form \cite{zhang}, which now possesses an SU(2) symmetry. Such a
dynamical realization of a spin-helix Hamiltonian constitutes an
intriguing consequence of the coupling between the mechanical mirror
oscillation and the effective spin-orbit interaction.

Results in Fig. 2 also indicate that in general cold-atom ZB
in a two-dimensional geometry suffers from quick damping
\cite{1Dcomment}, thus setting limitations to its potential
application. This ZB damping can have two different, but related,
interpretations. First, Eq. (\ref{com}) involves an integral over a
continuous distribution of the ZB frequency $\omega(\tilde{\bf p})$.
As time evolves, oscillations associated with different ${\bf p}$
comprising the initial wavepacket will necessarily dephase. This
makes it clear that the lifetime of the ZB is inversely proportional
to the momentum spread in the initial wavepacket
\cite{VaishnavPRL2008}. The second picture is more enlightening.
Note that the oscillating term in Eq. (\ref{com}) is due to the
quantum interference between the $|\psi^{+}\rangle$ and
$|\psi^{-}\rangle$ branches in Eq. (\ref{psit}). Because these two
terms have different group velocities ${\bf \nabla}_{\bf p}
E_{\pm}({\bf p},\tilde{\bf p})={\bf p}/{m} \pm  \hbar\kappa {\bf
\nabla}_{\bf p} |\tilde{\bf p}|/m$, the wavepacket on the
$|\psi^{+}\rangle$ branch will move away from that on the
$|\psi^{-}\rangle$ branch (see insets in Fig. 2). As a consequence the
overlap of the two wavepackets decreases with time, their quantum
interference decays, and hence ZB damps. The CDT condition
${\cal J}_0(\frac{2\kappa v_{d}}{\omega_d})=0$ may eliminate this
group velocity difference, but as mentioned above, it also kills ZB
in the beginning.

To dynamically suppress the ZB damping of a two-dimensional cold-atom wavepacket, we next consider a different regime of
$\omega_{d}$. Suppose the initial wavepacket has an average momentum
$p_{x}^0$ ($p_{z}^0$) along the $x$ ($z$) direction, with
$p_x^{0}\gg p_z^0$. We let $\omega_d$ satisfy the following
``resonance" condition: $p_{x}^{0} \approx m
\omega_{d}/2\kappa$, with $\omega_{d}\gg \kappa p_z^0/m, \kappa
v_{d}$. In this regime, $\omega_{d}$ matches the precession
frequency of the effective spin around $p_x^0$. However, in a
rotating frame, where the spinor wavefunction $(a', b')^{\text{T}}$
is related to the wavefunction $(a,b)^{\text{T}}$ in a non-rotating
frame by
\begin{equation}
 \left(\begin{array}{c} a'\\
b'
\end{array}
 \right)=\frac{a+b}{2}\left(\begin{array}{c} 1\\
1
\end{array}
 \right)e^{\frac{i\omega_{d}t}{2}}+\frac{a-b}{2}\left(\begin{array}{c} 1\\
-1
\end{array}
 \right)e^{-\frac{i\omega_{d}t}{2}},
\end{equation}
$\omega_{d}$ still dominates over all other frequencies. One can
then apply again the above-introduced high-frequency approximation
and obtain the following effective Hamiltonian in the rotating frame
(for $v_{d} \kappa/\omega_{d}\ll1$)
\begin{equation} \label{RWAHamiltonian1}
H^{\text{eff}}_{\text{reso}}=\frac{{\bf p}^{2}}{2m}+\frac{\hbar
\kappa}{m}\left[\frac{mv_{d}}{2}\sigma_z +
(p_{x}-\frac{m\omega_{d}}{2\kappa})\sigma_x\right].
\end{equation}
Eigenvalues of this effective on-resonance Hamiltonian are given by $
E_{\pm}^{\text{reso}}({\bf p}) = {\bf p}^{2}/2m \pm (\hbar\kappa/m) \left[ \left(mv_{d}/2\right)^2+ \left(p_x-m\omega_{d}/2\kappa\right)^2\right]^{1/2}
\approx {\bf p}^{2}/2m \pm \hbar \kappa v_{d}/2$, where we have used
the resonance condition $p_{x}^{0} \approx m \omega_{d}/2\kappa$ and the
assumption that the $p_x$ of a wavepacket is strongly peaked at $p_x^0$. Same
as in our early consideration, the beating frequency between the two
eigenvalue branches, i.e., $\omega^{\text{reso}}({\bf p})\equiv
E_{+}^{\text{reso}}({\bf p})/\hbar- E_{-}^{\text{reso}}({\bf
p})/\hbar$, gives rise to the new ZB frequency
$\omega^{\text{reso}}({\bf p})\approx \kappa v_{d}$. This indicates
that now the ZB frequency is totally determined by the velocity amplitude $v_d$ of the mirror oscillation,
thus entirely converting a mechanical property of the mirror to that of a
fundamental quantum coherence phenomenon. Further, it is easy to
find that $\langle {\bf r}\rangle$ here can still be given by Eq.
(\ref{com}), but with $\tilde{{\bf p}}=(p_x - mv_d /2\kappa)
\hat{e}_x+ (mv_d/2)\hat{e}_z\approx (mv_d/2) \hat{e}_z$. Further
evaluating ${\bf \nabla}_{\bf p}\theta(\tilde{\bf p})$ in Eq.
($\ref{com}$), one finds that the ZB here should be along the $x$
direction.

\begin{figure}[t]
\begin{center}
\vspace*{-2cm}
\par
\resizebox *{12cm}{10cm}{\includegraphics*{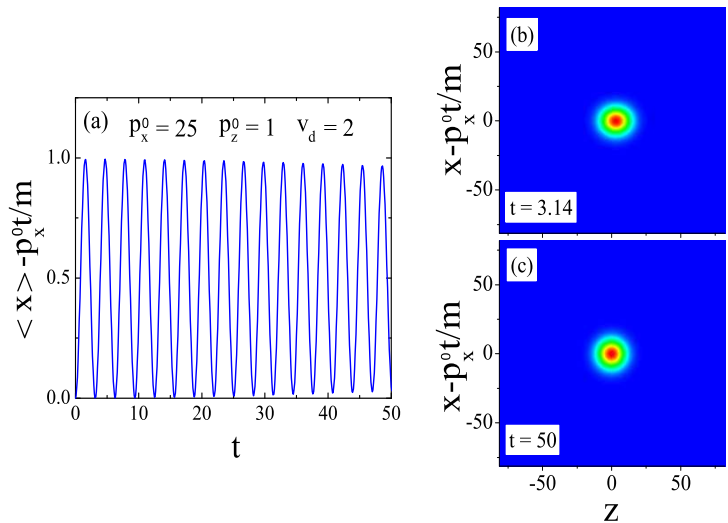}}
\end{center}
\par
\vspace*{-0.5cm} \caption{(Color Online) (a) Suppression of ZB
damping by mirror oscillation in an on-resonance regime. The initial
momentum spread and $\omega_d$ are the same as in Fig. 2. 
Panels (b) and (c) demonstrate that the wavepacket does not split despite the ZB.  }
\end{figure}

Significantly, the group velocities for both the eigenvalue branches
$E_{\pm}^{\text{reso}}({\bf p})$ now become the same, i.e.,
$\nabla_{\bf p} E_{\pm}^{\text{reso}}({\bf p}) \approx \frac{\bf
p}{m}$. Hence, an initial wavepacket undergoing ZB will not split
into two parts. According to our early explanation of the ZB damping
seen in Fig. 2, such suppression of wavepacket splitting should
suppress the damping of ZB. Numerical results in Fig. 3 directly
using $H_{\text{driven}}$ confirm our predictions based on
$H_{\text{reso}}^{\text{eff}}$. In particular, the initial state
used in Fig. 3 has the same momentum spread as in Fig. 2, and the
mirror oscillation is now under the resonance condition $p_{x}^{0}
\approx m \omega_{d}/2\kappa$. It is seen from Fig. 3 that the
wavepacket does not split despite the ZB. For a similar ZB frequency
as in Fig. 2(a), the ZB damping can be hardly seen in Fig. 3,
even after doubling the time scale. It can be estimated that via the
mirror oscillation the ZB lifetime here is increased by more than one
order of magnitude.

\section{CONCLUSION}

To conclude, by considering mirror oscillation in a laser-atom
system, we have proposed to explore time-dependent Dirac-like
equations via driving an effective spin-orbit interaction. Using
cold-atom ZB as a case study, we have shown how mirror oscillation
can be used to control the amplitude, the frequency, and the
lifetime of ZB in a two-dimensional geometry. For $m\sim10^{-25}$
Kg, $\kappa\sim 10^{6}$ m$^{-1}$, $|{\bf p}|\sim  1-10\ \hbar\kappa
$, we find
$\omega_{m} \sim 10^{4}$ Hz, falling in the range of the mirror oscillation frequency in current
optomechanical systems \cite{opo}. This also suggests that the peak
velocity of the mirror is of the order of $10^{-3}-10^{-2}$
ms$^{-1}$.
In addition to opening up a new means of matter-wave manipulation,
our theoretical work should greatly motivate experimental efforts in
realizing and exploring cold-atom Zitterbewegung.

{\bf ACKNOWLEDGMENTS}:
This work was supported by the CQT WBS grant No. R-710-000-008-271 (ZQ and
CH),  and by
the``YIA" fund (WBS grant No.: R-144-000-195-101) (JG) of the
National University of Singapore.

\end{document}